\documentclass[%
 reprint,
 superscriptaddress,
 showpacs,preprintnumbers,
 amsmath,amssymb,
 aps,
 prb,
]{revtex4-1}

\usepackage{graphicx}% Include figure files
\usepackage{dcolumn}% Align table columns on decimal point
\usepackage{bm}% bold math
\usepackage{upgreek}
\usepackage{amsmath}

\begin{document}

%\preprint{APS/123-QED}

\title{Polarization-controlled dynamically switchable high-harmonic generation from all-dielectric metasurfaces governed by dual bound states in the continuum}% Force line breaks with \\

\author{Shuyuan Xiao}
\email{syxiao@ncu.edu.cn}
\affiliation{Institute for Advanced Study, Nanchang University, Nanchang 330031, China}
\affiliation{Jiangxi Key Laboratory for Microscale Interdisciplinary Study, Nanchang University, Nanchang 330031, China}

\author{Meibao Qin}
\affiliation{School of Physics and Materials Science, Nanchang University, Nanchang 330031, China}

\author{Junyi Duan}
\affiliation{Institute for Advanced Study, Nanchang University, Nanchang 330031, China}
\affiliation{Jiangxi Key Laboratory for Microscale Interdisciplinary Study, Nanchang University, Nanchang 330031, China}

\author{Feng Wu}
\affiliation{School of Optoelectronic Engineering, Guangdong Polytechnic Normal University, Guangzhou 510665, China}

\author{Tingting Liu}
\email{ttliu@usst.edu.cn}
\affiliation{Institute of Photonic Chips, University of Shanghai for Science and Technology, Shanghai 200093, China}
\affiliation{Centre for Artificial-Intelligence Nanophotonics, School of Optical-Electrical and Computer Engineering, University of Shanghai for Science and Technology, Shanghai 200093, China}

\begin{abstract}
	
Tailoring optical nonlinear effects (e.g. harmonic generation, sum-frequency mixing, etc.) in the recently emerging all-dielectric platform is important for both the fundamental science and industrial development of high-efficiency, ultrafast, and miniaturized photonic devices. In this work, we propose a novel paradigm for dynamically switchable high-harmonic generation in Silicon nanodimer metasurfaces by exploiting polarization-controlled dual bound states in the continuum (BIC). Owing to the high-quality factor of BIC resonances, efficient harmonic signals including the third-harmonic generation and fifth-harmonic generation from a direct process as well as a cascaded process by degenerate four-wave mixing are obtained. Moreover, the BICs and their resonantly enhanced harmonics can be switched on or off with high selectivity respect to the fundamental pump polarization. Compared with previous reports, our work provide a simple but effective tuning strategy by fully exploring the structural symmetry and polarization degree of freedom rather than resorting to additional external stimuli, which would have great advantages in smart designing tunable and switchable nonlinear light source for chip-scale applications.

\end{abstract}

%\pacs{42.70.-a, 42.79.-e, 78.67.Pt}% PACS, the Physics and Astronomy
\maketitle

%\tableofcontents

\section{\label{sec1}Introduction}

Nonlinear optical processes such as high-harmonic generation have attracted vast interests in scientific community, which enjoy widespread applications ranging from lasing to imaging. The efficient conversion of fundamental pump to harmonic signal at the nanoscale is a key enabler for practical use\cite{Kauranen2012, Smirnova2016, Li2017}. Two main paths include the uses of highly nonlinear materials and high quality factor ($Q$ factor) resonant structures. The earlier approaches are based on plasmonic metasurfaces, but the disadvantage of intrinsic dissipative losses hinders their broad applicability\cite{Panoiu2018, Rahimi2018}. Alternatively, all-dielectric metasurfaces prove to be a promising platform for nonlinear optics\cite{Kuznetsov2016, Sain2019, Grinblat2021}. The high-index dielectric materials exhibit a wide spectrum of optical nonlinearities and tensor symmetries as well as negligible losses and high damage threshold, and the coexistence of electric and magnetic Mie resonances brings many intriguing effects in the nonlinear regime\cite{Carletti2015, Grinblat2016, Shorokhov2016, Xu2018, Gao2018, Frizyuk2019, Xu2020, Kroychuk2020, Yao2020}. In particular, a novel design strategy based on the concept of $bound$ $states$ $in$ $the$ $continuum$ (BIC) enables high-$Q$ resonances in all-dielectric metasurfaces, where the tight confinement of light inside the meta-atoms allows for efficient nonlinear frequency conversion\cite{Hsu2016, Koshelev2019a, Koshelev2019b}. To date, all-dielectric metasurfaces governed by BIC physics have demonstrated their ability to substantially boost second-harmonic generation (SHG)\cite{Vabishchevich2018, Carletti2018, Koshelev2020, Volkovskaya2020, Anthur2020, Ning2021, Huang2021}, and third-harmonic generation (THG)\cite{Yang2015, Tong2016, Xu2019, Koshelev2019, Carletti2019, Liu2019, Gandolfi2021}, with efficiency higher and pump intensity lower than previously recorded values. In a most recent report, empowered by the BIC resonance in asymmetric silicon metasurfaces, efficient harmonics up to the 11$^{\text{th}}$ order has been observed\cite{Zograf2022}.

The research agenda is now shifting towards realizing tunable and switchable functionalities to meet requirements of practical devices\cite{Zheludev2012, Xiao2020, Carletti2021}. Due to the strong dependence of optical responses in the nonlinear regime on those in the linear regime, effects on the harmonic wavelength can be precisely evaluated considering the perturbation on the fundamental wavelength, which lays the direct foundation for dynamical control of harmonic generation. Inspired by the recent progress of active metaoptics, resorting to external stimuli becomes a natural consideration and inevitable choice. For instance, the thermo-optical control of SHG in monolithic nanoantennas has been proposed via thermally tuning the material refractive index\cite{Celebrano2021, Rocco2021, Pashina2022}, and further extended to the ultrafast all-optical modulation via photoinjection of free carriers\cite{Pogna2021}. On the other hand, the indirect approach through integrating active media provides additional degree of freedom, though at the cost of sophisticated experimental requirements. The electro-optical material nematic liquid crystal\cite{Rocco2020}, and phase-change material chalcogenide glass\cite{Cao2019, Yue2021, Zhu2021, Abdelraouf2021, Liu2021} have been integrated with metasurfaces to change the surrounding dielectric environment for wavelength-tunable harmonic generation. Nevertheless, the introduction of external stimuli (regardless of the form) would increase the structural complexity and energy consumption. More seriously, the light confinement (evaluated by $Q$ factor) and consequently the harmonic conversion efficiency could not maintain the maximum due to the change of resonant conditions.

In this work, we propose a novel paradigm for dynamically switchable high-harmonic generation by fully exploiting polarization-controlled dual BIC resonances in all-dielectric nanodimer metasurfaces. In our design, two eigenmodes corresponding to symmetric-protected and accidental BICs can be selectively excited at two different fundamental wavelengths under two orthogonal polarizations, respectively. Benefiting from the ultrahigh $Q$ factor of BIC resonances, remarkably high conversion efficiencies of THG and fifth-harmonic generation (FHG) including both direct and cascaded processes are numerically demonstrated. By simply rotating the fundamental pump polarization, these two BIC resonances and their resonantly enhanced harmonics can be precisely switched on or off. In sharp contrast to above-mentioned approaches such as thermal tuning or utilizing active media, the present work affords an entirely different tuning strategy free from additional external stimuli, showing great prospects in designing metasurface platform for nonlinear devices and applications.

\section{\label{sec2}Polarization-controlled dual bound states in the continuum}

\begin{figure}[htbp]
\centering
\includegraphics% Here is how to import EPS art
[scale=0.37]{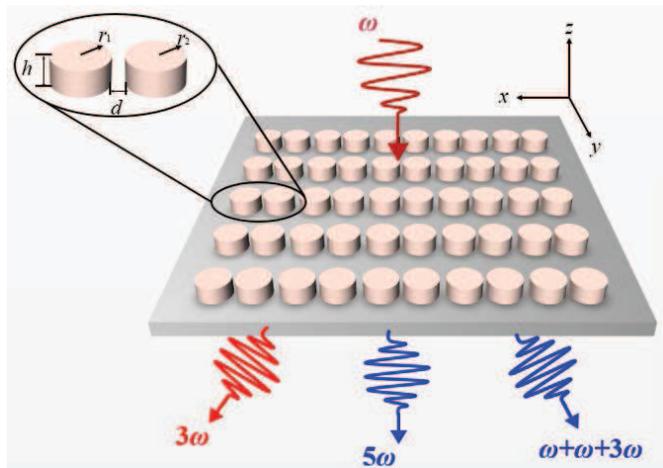}
\caption{\label{fig1} The schematic illustration of high-harmonic generation from the all-dielectric metasurfaces supporting dual BIC resonances. Si nanodimers are placed on top of SiO$_{2}$ substrate. Photons of a fundamental pump at the frequency $\omega$ are converted to photons at $3\omega$ via THG, and to photons at $5\omega$ via FHG and cascaded process of THG followed by degenerate four-wave mixing between the fundamental and THG signal. Top inset shows a unit cell of the metasurfaces. Geometric parameters: the period is $p=1020$ nm, the height is $h=300$ nm, and the radius and distance are initially set to $r_{1}=r_{2}=225$ nm and $d=40$ nm, respectively.}
\end{figure}

The example of the metasurfaces that we consider is schematically illustrated in Fig. \ref{fig1}. It consists of a periodic array of high-index dielectric nanodimers arranged in a square lattice on top of low-index substrate. There are many choices for material selection, in principle, depending on the wavelength of interest. Here the combination of silicon (Si) on silica (SiO$_{2}$) is adopted to generate odd optical harmonics (3rd and 5th) spanning the visible to ultraviolet. The wavelength-dependent complex refractive index of Si is extracted from experimental data\cite{Palik1985}, and the constant refractive index of 1.45 is assigned to SiO$_{2}$. Note that we choose the dimer design as it belongs to a symmetric geometry of $C_{2v}$ group and possesses more structural parameters to be tuned than $C_{4v}$ configurations\cite{Rocco2018, He2018, Yu2019, Song2019, Zhou2020, Overvig2020, Song2020, Frizyuk2021, Parry2021, Xu2022}. Benefiting from the additional degree of freedom, it is possible to manipulate the excitation of dual BIC resonances by adjusting the structural geometry and rotating the fundamental pump polarization. Under normal incidence, we consider the transmitted high-harmonic signals including the conventional THG and FHG from a direct process as well as FHG via a cascaded process of THG followed by degenerate four-wave mixing between the fundamental and THG signal. 

\begin{figure}[htbp]
\centering
\includegraphics% Here is how to import EPS art
[scale=0.38]{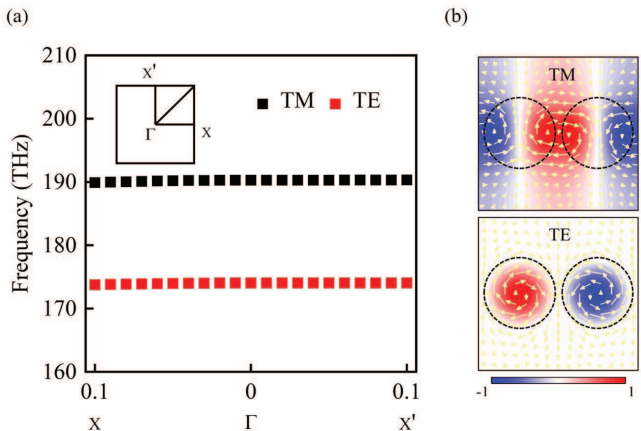}
\caption{\label{fig2} (a) The band diagram along $X$-$\Gamma$-$X'$ ($kp/2\pi$) of Si nanodimer metasurfaces. (b) The corresponding electromagnetic field distributions at $\Gamma$ point for the TM-like and TE-like modes, respectively. Top: the $z$-component of electric field $E_{z}$ overlaid with the in-plane magnetic field direction vector $H_{xy}$. Bottom: the $z$-component of magnetic field $H_{z}$ overlaid with the in-plane electric field direction vector $E_{xy}$.}
\end{figure}

Considering that the excitations of BIC resonances at the fundamental frequency predict the efficient harmonic generation, we perform the eigenmode analysis to characterize and optimize the modal properties of the Si nanodimer metasurfaces. The eigenmodes are calculated with the eigenfrequency solver in COMSOL Multiphysics with Floquet periodic boundary conditions applied in the $x$ and $y$ directions and perfectly matched layers used in the $z$ direction. In Fig. \ref{fig2}(a), the band diagram along $X$-$\Gamma$-$X'$ are presented consisting of a transverse magnetic (TM)-like mode and a transverse electric (TE)-like mode in the first brillouin zone. For the TM eigenmode, the complex eigenfrequency is $\widetilde v=190.29$ THz containing no imaginary part, and thus the $Q$ factor (defined as $Q=\frac{1}{2}\frac{\text{Re}(\widetilde v)}{\text{Im}(\widetilde v)}$) diverges towards infinity, which indicates it is a BIC with no leakage coupled to the radiation continuum; for the TE eigenmode, $\widetilde v=174.04+0.0084$ THz, and the $Q$ factor is larger than $10^{4}$, which suggests it is a quasi BIC that may be transformed into a BIC after parameter tuning. Resorting to the $C_{2}$ group symmetry analysis can help us understand these two BICs in a more intuitive way. Figure \ref{fig2}(b) presents the electromagnetic field distributions at $\Gamma$ point (the particular high-symmetry point of the reciprocal space), the $z$-component of electric field $E_{z}$ of the TM eigenmode is even under 180$^{\circ}$ rotation around the incident direction, i.e. $z$ axis, while the $z$-component of magnetic field $H_{z}$ of the TE eigenmode is odd. Given that the plane wave propagating along the $z$ axis is the only radiating channel at $\Gamma$ point for the subdiffractive regime and its electromagnetic fields are odd under the same rotation\cite{Hsu2016}, the TM eigenmode can not be excited by a normally incident plane waves with any linear polarization due to symmetry mismatch (except when the structural symmetry breaking is presented which will be discussed in the following). The TE eigenmode, in sharp contrast, can be excited by a $y$-polarized normally incident plane wave.

\begin{figure}[htbp]
\centering
\includegraphics% Here is how to import EPS art
[scale=0.40]{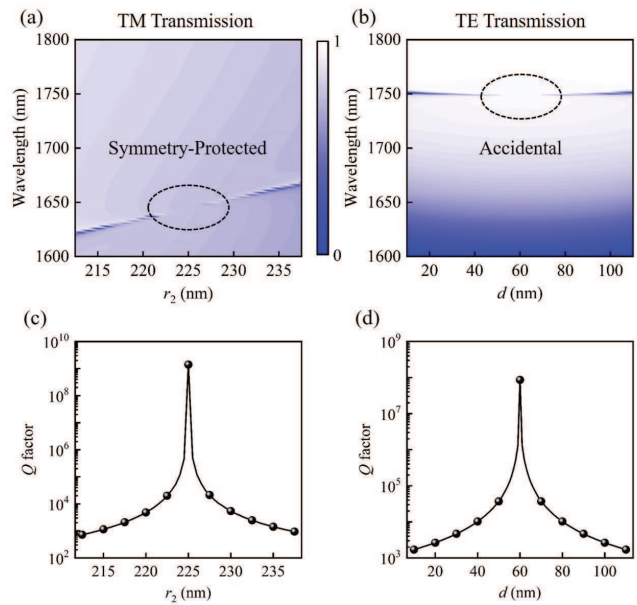}
\caption{\label{fig3} The transmission spectra of Si nanodimer metasurfaces as a function of (a) the radius $r_{2}$ and (b) the distance $d$. (c) and (d) The corresponding $Q$ factors for the resonances in (a) and (b), respectively. The $x$- and $y$-polarized plane wave is normally incident to the metasurfaces for the numerical calculations of (a), (c), and (b), (d), respectively.}
\end{figure}

Then we provide a visual insight using the transmission spectra of the Si nanodimer metasurfaces in Figs. \ref{fig3}(a) and \ref{fig3}(b). The frequency domain solver, instead of the eigenfrequency, is employed for calculation with two ports placed at the interior boundaries of the perfectly matched layers for both setting an excitation plane wave source and recording
the transmission in terms of scattering $S$-parameters. Under $x$-polarized normal incidence, the resonance (identified through the transmission dip close to 0) is invisible when the structural symmetry is presented, i.e., the radii of the nanodimers are the same $r_{1}=r_{2}=225$ nm, which corresponds to the TM eigenmode. When $r_{2}$ deviates from $r_{1}$, the resonance emerges and becomes pronounced in the wavelength range of 1600–1700 nm. The symmetry breaking via changing $r_{2}$ opens a leaky channel coupled to the continuum and transforms BIC to quasi BIC with radiation loss, validating the TM eigenmode as a symmetry-protected BIC. We also calculate the $Q$ factors with the eigenfrequency solver. In Fig. \ref{fig3}(c), the $Q$ factor of the TM eigenmode experiences a dramatic increase as $|r_{1}-r_{2}|$ reduces, and diverges at the BIC region near $r_{1}=r_{2}$ with infinitely large value ($>10^{9}$). On the other hand, the leaky resonance associated with the TE eigenmode under $y$-polarized excitation can be easily found around the wavelength of 1750 nm. Its radiation channel can be closed by changing the spacing distance $d$ of the nanodimers, rendering infinitely small resonance width at $d=60$ nm, which is a typical parameter tuning behavior of the so-called accidental BIC. In contrast to the symmetry-protected BIC, there is no symmetry perturbation in the parameter tuning of the accidental BIC\cite{Hsu2016, He2018, Zhou2020, Xu2022}. In Fig. \ref{fig3}(d), the corresponding $Q$ factor reaches the peak value tending to be infinite large ($>10^{8}$) for $d=60$ nm. 

\section{\label{sec3}Nonlinear harmonic generation}

To investigate the potential of the dual BIC in high-harmonic generation, we perform the nonlinear simulation implemented with the frequency domain solver in COMSOL Multiphysics. We assume the undepleted pump approximation and employ two coupled electromagnetic models in consecutive steps. First, the transmission at the fundamental wavelength is simulated to retrieve the local field distributions and compute the nonlinear polarization induced inside the structure. Second, this nonlinear polarization is employed as the only source for the next simulation at the harmonic wavelength to obtain the generated field and recover the nonlinear power flux radiated to the substrate. 

\subsection{\label{sec3.1}Third-harmonic generation}

\begin{figure}[htbp]
\centering
\includegraphics% Here is how to import EPS art
[scale=0.18]{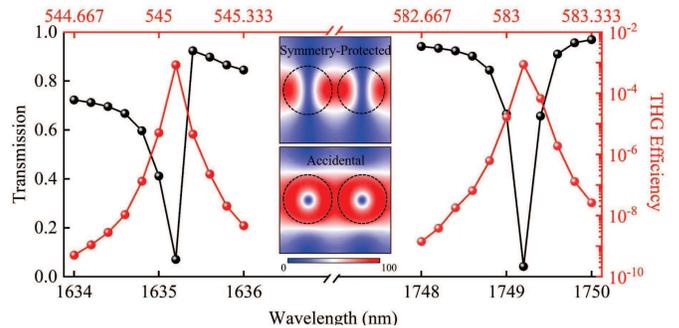}
\caption{\label{fig4} The transmission and THG efficiency spectra of Si nanodimer metasurfaces as a function of the pump wavelength near the BIC resonances. (left panel) Symmetry-protected BIC: $r_{1}=225$ nm, $r_{2}=220$ nm, and $d=40$ nm. (right panel) Accidental BIC: $r_{1}=225$ nm, $r_{2}=225$ nm, and $d=40$ nm. The $x$($y$)-polarized plane wave is normally incident to the metasurfaces for the numerical calculations of the left (right) panel. Middle insets show the in-plane electric field distributions at the wavelength of symmetry-protected and accidental BIC resonance, respectively.}
\end{figure}

\begin{figure}[htbp]
\centering
\includegraphics% Here is how to import EPS art
[scale=0.40]{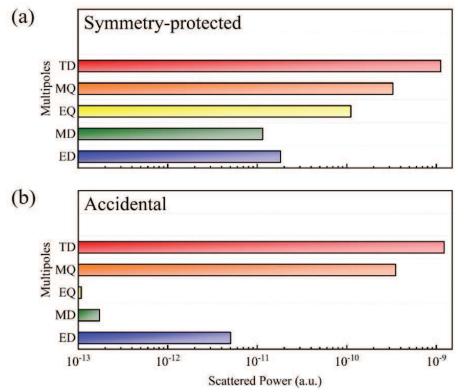}
\caption{\label{fig5} Contributions of multipole moments in the Cartesian coordinates, including the electric dipole (ED), magnetic dipole (MD), electric quadrupole (EQ), magnetic quadrupole (MQ), and toroidal dipole (TD) to the far-field radiation of (a) symmetry-protected and (b) accidental BIC resonance.}
\end{figure}

We first study the THG from the proposed Si nanodimer metasurfaces. The nonlinear polarization can be calculated by the local electric field at the fundamental wavelength, i.e., $\bm{P}^{3\omega}=3\varepsilon_{0}\chi^{(3)}(\bm{E\cdot\bm{E}})\bm{E}^{\omega}$, where $\varepsilon_{0}=8.8542\times10^{-12}$ F/m is the vacuum permittivity, and $\chi^{(3)}=2.45\times10^{-19}$ m$^{2}$/V$^{2}$ is the third-order nonlinear susceptibility of Si in the near-infrared. The conversion efficiency of THG is defined as $\eta_{\text{THG}}=\frac{P_{\text{THG}}}{P_{\text{Pump}}}$, where $P_{\text{THG}}$ and $P_{\text{Pump}}$ are the THG and Pump powers, respectively. As above mentioned, the symmetry-protected and accidental BIC that we are considering here are excited with the $x$- and $y$-polarized plane wave, respectively, therefore we explore these two orthogonal polarization scenarios to selectively enhance optical coupling from the fundamental pump. With the optimized geometric parameters derived from Fig. \ref{fig2}, the simulated transmission and THG efficiency spectra are presented in Fig. \ref{fig4}. Under $x$-polarized incidence (left panel), the symmetry-protected BIC transforms into a high-$Q$ resonance located at 1635.2 nm when the structural symmetry becomes broken ($r_{1}=225$ nm and $r_{2}=220$ nm). Under $y$-polarized incidence (right panel), a leaky resonance stemming from the accidental BIC is observed at 1749.2 nm after the distance tuning ($d=40$ nm). We further perform the Cartesian multipolar decomposition to quantitatively identify the multipole characters of the high-$Q$ BIC resonances. In Figs. \ref{fig5}(a) and \ref{fig5}(b), it is observed that the dominant contribution at resonance is provided by the toroidal dipole (TD) in both cases. Here it radiates stronger than the magnetic quadrupole (MQ) by a factor of $\sim 3.5$, and stronger than the radiating component of the electric dipole (ED), magnetic dipole (MD), and electric quadrupole (EQ) by orders of magnitude. The TD moments can trap the incident light with the circular head-to-tail dipole arrangement (which is exactly consistent with the field direction vectors in Fig. \ref{fig2}b) and substantially enhance the local field. To gain a deeper impression, the corresponding electric field distributions at the wavelength of 1635.2 nm and 1749.2 nm are depicted in the insets of Fig. \ref{fig4}. Significant enhancement up to 100 fold is presented in the vicinity of both BIC resonances, leading to the substantial increase of the THG conversion efficiency as high as $10^{-3}$ at an incident excitation intensity of 1 MW/cm$^{2}$. Accompanying the narrow transmission spectra, the linewidth of THG efficiency spectra is quite small with the full-width at half-maximum (FWHM) less than 1 nm, exhibiting good monochromaticity due to the resonantly enhanced mechanism. It is worth pointing out that the two intense THG signals are emitted at different harmonic wavelengths under orthogonal polarizations, hence the occurrence of the nonlinear signals has no mutual influence between them, which implies the possibility of independent control. 

\begin{figure}[htbp]
\centering
\includegraphics% Here is how to import EPS art
[scale=0.40]{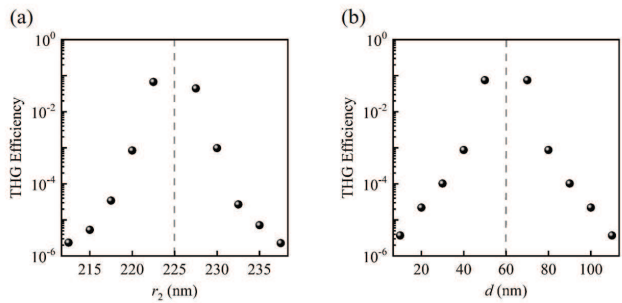}
\caption{\label{fig6} The THG efficiencies of Si nanodimer metasurfaces as a function of (a) the radius $r_{2}$ and (b) the distance $d$. The $x$- and $y$-polarized plane wave is normally incident to the metasurfaces for the numerical calculations of (a) and (b), respectively.}
\end{figure}

The asset of our work is the use of the dual BIC resonances both with ultrahigh $Q$ factors. To investigate the impact of $Q$ factor on the THG, we provide the modification of conversion efficiencies when the geometry parameters are detuned from the perfect BIC configurations under $x$- and $y$-polarized incidence in Figs. \ref{fig6}(a) and \ref{fig6}(b), respectively. The THG conversion efficiencies show similar dependence on the radius $r_{2}$ and spacing distance $d$, in comparison with the variations of $Q$ factors in Figs. \ref{fig3}(c) and \ref{fig3}(d). The THG efficiency increases rapidly from $10^{-6}$ to $10^{-1}$ as $r_{2}$ gradually approaches 225 nm of the symmetry-protected BIC region, or with $d$ closes to 60 nm of the accidental BIC region. This can be explained by the higher $Q$ factors and in turn larger field enhancement when the geometric parameters become closely matched to those of the BIC. Note that the distributions of THG efficiency dependent on $d$ in Fig. \ref{fig6}(b) is symmetric on the two sides of $d=60$ nm, resembling the $Q$ factor distribution in Fig. \ref{fig3}(d). The reason behind this is the intriguing geometry design of nanodimer metasurfaces. At the very special point at $d=60$ nm, the distance between centers of the nanodimers equals to half of the period. The increasing distance with $+\Delta d$ will separate the nanodimers in the original unit cell and combine them with the neighbouring partners forming new unit cells, leading to the same collective effect with the case $-\Delta d$. 

\subsection{\label{sec3.2}Fifth-harmonic generation}

\begin{figure}[htbp]
\centering
\includegraphics% Here is how to import EPS art
[scale=0.19]{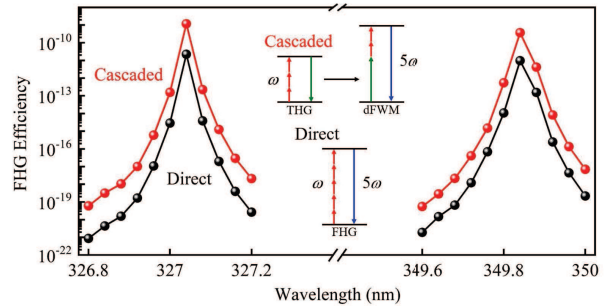}
\caption{\label{fig7} The FHG efficiency spectra of Si nanodimer metasurfaces as a function of the pump wavelength near the BIC resonances. (left panel) Symmetry-protected BIC: $r_{1}=225$ nm, $r_{2}=220$ nm, and $d=40$ nm. (right panel) Accidental BIC: $r_{1}=225$ nm, $r_{2}=225$ nm, and $d=40$ nm. The $x$($y$)-polarized plane wave is normally incident to the metasurfaces for the numerical calculations of the left (right) panel. Middle insets show the schemes of the direct and cascaded processes of FHG.}
\end{figure}

The ultrahigh $Q$ factor and intense electric field enhancement supported by the dual BIC resonances are advantageous to the high-order harmonic generation. This field has been dominated by gas-based systems, suffering from expensive vacuum steps and complicated methods, which can be relieved using the solid state physics platform\cite{Sivis2017, Vampa2017, Liu2020, Shcherbakov2021}. Here we consider the fifth harmonic signals from the proposed Si nanodimer metasurfaces, including two of potential nonlinear processes, as depicted in the inset of Fig. \ref{fig7}. One common path to obtain efficient FHG is using a direct process, i.e., $\bm{P}^{5\omega}=3\varepsilon_{0}\chi^{(5)}(\bm{E\cdot\bm{E}})^{2}\bm{E}^{\omega}$, where $\chi^{(5)}$ is the fifth-order nonlinear susceptibility of Si. In the numerical simulations of direct FHG, $\chi^{(5)}$ is estimated as $5.6\times10^{-39}$ m$^{4}$/V$^{4}$. At an incident excitation intensity of 1 MW/cm$^{2}$, the direct FHG efficiency spectra as a function of the wavelength in the vicinity of dual BIC resonances are depicted in Fig. \ref{fig7}. The peak conversion efficiency of $10^{-11}$ is achieved at respective harmonic wavelength, with one assisted by the high-$Q$ resonance from the symmetry-protected BIC under $x$-polarized incidence and the other owing to the accidental BIC under $y$-polarized incidence. 

On the other hand, cascaded nonlinear process provides an alternative path to high harmonic generation. We consider the FHG using a cascaded $\chi^{(3)}$ effect consisting of THG followed by degenerate four-wave mixing between the fundamental and THG signal, as shown in Fig. \ref{fig1}. In the mixing process, the fundamental pump of frequency $\omega$ is referred to as the pump and the TH signal of frequency $3\omega$ as the idler. With the two input frequencies, the frequency of the generated signal photon is denoted by the sum of the input frequencies as $5\omega=\omega+\omega+3\omega$. Hence the nonlinear polarization of the degenerate four-wave mixing process can be defined as $\bm{P}^{5\omega}=3\varepsilon_{0}\chi^{(3)}[2(\bm{E^{\omega}\cdot\bm{E}^{3\omega}})\bm{E}^{\omega}+(\bm{E}^{\omega}\cdot\bm{E}^{\omega})\bm{E}^{3\omega}]$. For comparison, the optical intensity of fundamental pump is fixed to 1 MW/cm$^{2}$ during the simulations. The calculated FHG efficiency achieves a peak value of $10^{-9}$ under both polarizations. The efficiencies are higher by approximately two orders of magnitude in the cascaded process compared to that obtained in the direct case. Considering the same geometric parameters and resonance modes of the two cases, this interesting phenomenon can be attributed to the relation between $\chi^{(5)}$ and $\chi^{(3)}$, i.e., $\chi^{(5)}\approx[\chi^{(3)}]^{2}/\chi^{(1)}$.

\subsection{\label{sec3.3}Polarization control of harmonic generation}

As mentioned previously, the dual BIC resonances and resonantly enhanced harmonic signals from the proposed Si nanodimer metasurfaces are highly dependent on the polarization property of the incident light, making it possible to realize the polarization control of harmonic generation. In the final section, we will demonstrate this with a simple example. The geometric parameters are fixed to $r_{1}=225$ nm, $r_{2}=220$ nm, and $d=40$ nm so that the high-$Q$ resonances arising from the symmetry-protected and accidental BIC can be well supported at their respective fundamental wavelengths under two orthogonal polarizations.

\begin{figure}[htbp]
\centering
\includegraphics% Here is how to import EPS art
[scale=0.40]{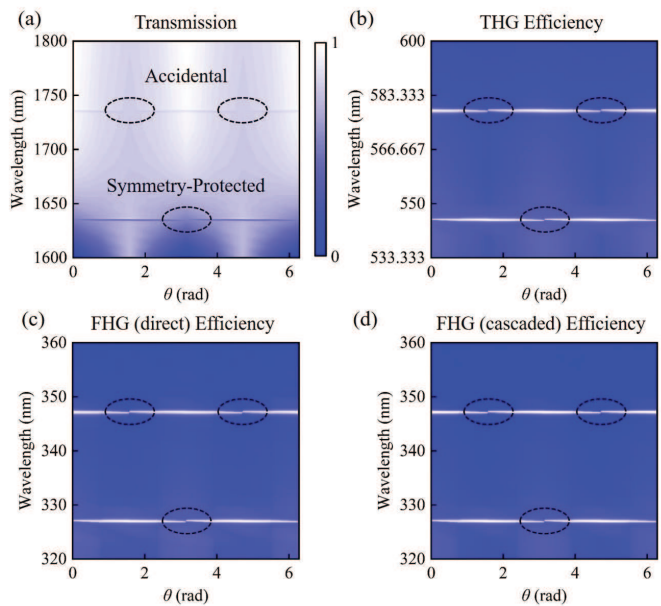}
\caption{\label{fig8} (a) The transmission and (b)-(d) THG, FHG (direct), FHG (cascaded) efficiency spectra of Si nanodimer metasurfaces as a function of the polarization angle of the incident plane wave. Geometric parameters: $r_{1}=225$ nm, $r_{2}=225$ nm, and $d=40$ nm.}
\end{figure}

Figure \ref{fig8}(a) presents the transmission spectra of the Si nanodimer metasurfaces as a function of the polarization angle, with two resonances clearly identified as the symmetry-protected BIC at 1635.2 nm and the accidental BIC at 1736 nm, respectively. The resonant wavelength of the latter shows a slight blueshift from 1749.2 nm since $r_{2}$ deviates from $r_{1}$ to meet the excitation condition of the former. It is very important to point out that both BIC resonances are not completely continuous with respect to the evolution of the polarization angle, specifically, the accidental BIC is absent at $\theta=\frac{1}{2}\pi$ and $\frac{3}{2}\pi$, and the symmetry-protected BIC is absent at $\theta=0$, $\pi$, and $2\pi$, which correspond to the $x$- and $y$-polarization in our setup, respectively. This phenomenon can be well explained by the dimer design governed by the symmetric geometry of $C_{2v}$ group. Under such circumstance, the dual BIC resonances can be selectively excited at two different fundamental wavelengths under different polarizations. The most important consequence, in an intuitive way, is that the BIC resonantly enhanced THG and FHG signals can be switched on or off by simply rotating the fundamental pump polarization. As shown in Figs. \ref{fig8}(b), \ref{fig8}(c), and \ref{fig8}(d), we have seen the predicted functionalities. We have thus realized dynamically switchable high-harmonic generation without any additional external stimuli: when $\theta$ equals odd multiples of $\frac{1}{2}\pi$, only the harmonics pumped from 1635.2 nm are emitted, and when $\theta$ equals even multiples of $\frac{1}{2}\pi$, only the harmonics pumped from 1736 nm are emitted, in all other cases both signals are emitted. Worth mentioning at this time is the maximum efficiency which is maintained regardless of the polarization angle by complete utilization of the pump light.

\section{\label{sec4}Conclusions}

In conclusion, we design the polarization-controlled dual BIC resonances in Si nanodimer metasurfaces, and apply them as a solution to dynamically switchable harmonic generation by utilizing the polarization dependence. Due to the high $Q$ features of BIC resonances, hihg-efficiency harmonic signals pumped from the infrared to the visible through THG ($10^{-3}$) and the ultraviolet through FHG including both the direct ($10^{-11}$) and cascaded ($10^{-9}$) processes are obtained at a moderate optical intensity of 1 MW/cm$^{2}$. By simply rotating the fundamental pump polarization, the BIC resonances and their resonantly enhanced harmonics can be selectively switched on or off without the aid of additional external stimuli. The proposed paradigm should have great merits in designing the next-generation nonlinear light source for chip-scale applications. Finally, it is worth pointing out that the strategy and structure here are rather general, and can be extended for the selective enhancement of other nonlinear effects at alternative wavelengths via physical scalability\cite{Joannopoulos2008, Koshelev2018}. One example that comes to mind may be the polarization-controlled harmonic generation and sum-frequency mixing in 2D materials\cite{Lin2019, Yuan2019, Bernhardt2020, Loechner2020, Liu2021a}.

\begin{acknowledgments}	
	
This work is supported by the National Natural Science Foundation of China (Grants No. 11947065, No. 61901164, and No. 12104105), the Natural Science Foundation of Jiangxi Province (Grant No. 20202BAB211007), the Interdisciplinary Innovation Fund of Nanchang University (Grant No. 2019-9166-27060003), the Start-up Funding of Guangdong Polytechnic Normal University (Grant No. 2021SDKYA033), and the China Scholarship Council (Grant No. 202008420045). The authors would like to thank T. Ning, L. Huang, and T. Guo for beneficial discussions on the nonlinear numerical simulations.

S. X. and M.Q. contributed equally to this work.

\end{acknowledgments}

%\bibliography{Ref}% Produces the bibliography via BibTeX.


\end{document}